\documentclass[a4paper,10pt,twoside]{cpc-hepnp}

\usepackage{multicol}
\usepackage{graphicx}
\usepackage{booktabs}
\usepackage{amssymb,bm,mathrsfs,bbm,amscd}
\usepackage[tbtags]{amsmath}
\usepackage{lastpage}
\usepackage{hyperref}

\begin{document}

\fancyhead[c]{\small "Submitted to 'Chinese Physics C'"}
\fancyfoot[C]{\small \thepage}

\title{Method Study of Parameter Choice for a Circular Proton-Proton Collider \thanks{Supported by National Natural Science
Foundation of China (11175192) }}

\author{%
      SU Feng£¨Ëշ壩$^{1}$\email{sufeng@ihep.ac.cn}%
\quad GAO Jie£¨¸ß½Ü£©
\quad XIAO Ming£¨Ð¤Ãú£©
\quad WANG Dou£¨Íõ¶º£©\\
\quad WANG Yi-Wei£¨ÍõÒãࣩ
\quad BAI Sha£¨°×ɯ£©
\quad BIAN Tian-Jian£¨±ßÌì½££©
}
\maketitle

\address{%
Key Laboratory of Particle Acceleration Physics and Technology, Institute of High Energy Physics,Chinese Academy of Sciences, Beijing 100049, China
}

\begin{abstract}
In this paper we showed a systematic method of appropriate parameter choice for a circular pp collider by using analytical expression of beam-beam tune shift limit started from given design goal and technical limitations. A parameter space has been explored. Based on parameters scan and considerations from RF systems, a set of appropriate parameter designed for a 50Km and a 100Km circular proton-proton collider was proposed.
\end{abstract}



\footnotetext[0]{\hspace*{-3mm}\raisebox{0.3ex}{$\scriptstyle\copyright$}2015
Chinese Physical Society and the Institute of High Energy Physics
of the Chinese Academy of Sciences and the Institute
of Modern Physics of the Chinese Academy of Sciences and IOP Publishing Ltd}%

\begin{multicols}{2}

\section{Introduction}

With the discovery of Higgs boson on LHC, the world high-energy physics community is investigating the feasibility of a Higgs Factory as a complement to the LHC for studying the Higgs and interested in the frontier of high energy. The CERN people are busy planning the LHC upgrade program, including HL-LHC and HE-LHC. They also plan a more inspiring program called FCC, including FCC-ee and FCC-hh. Both the HE-LHC and the FCC-hh are proton-proton colliders aiming to explore the high energy frontier and expecting to find new physics \citep{lab1}\citep{lab2}\citep{lab3}\citep{lab4}. Chinese accelerator physicists also plan to design an ambitious machine called CEPC-SPPC(Circular Electron Positron Collider-Super Proton Proton Collider). The CEPC-SPPC program contains two stage. The first stage is an electron-positron collider with center-of-mass energy 240GeV to study Higgs properties carefully. The second stage is a proton-proton collider at center-of-mass energy more than 70TeV \citep{lab5}\citep{lab6}\citep{lab7}. The SPPC design is just starting. We developed a systematic method of how to make an appropriate parameter choice for a circular pp collider by using analytical expression of beam-beam tune shift started from the required luminosity goal, beam energy, physical constraints at IP and some technical limitations.

\section{Beam-Beam tune shift limit}

In storage ring colliders, due to quantum excitation and synchrotron damping effects, the particles are confined inside a bunch. In $e^+ e^-$ colliders, the quantum excitation is very strong and the position for each particle is random and the state of the particles can be regarded as a gas, where the positions of the particles follow statistic laws. Apparently, the synchrotron radiation is the main source of heating. Besides, when two bunches undergo collision at an interaction point (IP), every particle in each bunch will feel the deflected electromagnetic field of the opposite bunch and the particles will suffer from additional heatings. With the increase of the bunch particle population $N_e$, this kind of heating effect will get stronger. There is a limit condition beyond which the beam emittance will blow up. This emittance blow-up mechanism introduce a limit for beam-beam tune shift which was well discussed in reference \citep{lab8}:

\begin{equation}
\xi_{y,max}\leq 2845\gamma\sqrt{\frac{r_p}{6\pi R N_{IP}}}=\frac{2845}{2\pi}\sqrt{\frac{T_0}{\tau_y \gamma N_{IP}}}
\end{equation}

In $pp$ circular colliders, the synchrotron damping effect is very weak. The position for each particle is not like that for electron which is random and the state of the particles cannot be regarded as a gas. Due to the lack of strong synchrotron radiation, the particles inside a bunch are very cold and one can trace each particle without missing it. When the bunches suffer from the strong nonlinear beam-beam forces, some particles located in the outer part of the bunch undergo nonlinear force induced stochastically motions. The number of this heated particles, $N_{p,h}$ can be estimated by $N_{p,h}=f(x)N_{p}$ \citep{lab9}. With

\begin{equation}
f(x) = 1- \frac{2}{ \sqrt{2\pi}}\int_0^x \mathrm{e}^{{-}\frac{t^2}{2}} \,\mathrm{d}t
\end{equation}

Where $N_{p}$ is the particle number inside a bunch, $x$ is the limit between the cold core and the heated region. On this condition, the limit for beam-beam tune shift can be expressed as \citep{lab9}:

\begin{equation}
\xi_{y,max}= \frac{2845\gamma}{f(x)}\sqrt{\frac{r_p}{6\pi R N_{IP}}}=\frac{2845}{2\pi f(x)}\sqrt{\frac{T_0}{\tau_y \gamma N_{IP}}}=\frac{\xi_1}{f(x)}
\end{equation}

\begin{equation}
f(x) = 1- \frac{2}{ \sqrt{2\pi}}\int_0^x \mathrm{e}^{{-}\frac{t^2}{2}} \,\mathrm{d}t
\end{equation}

\begin{equation}
x^2=\frac{4f(x)}{\pi \xi_{y,max} N_{IP}}=\frac{4f(x)^2}{\pi \xi_1 N_{IP}}
\end{equation}

Where $N_{IP}$ is the number of interaction point (When there are $N_{IP}$ interaction points, the independent heating effects have to be added in a statistical way), $R$ is the dipole radius, $r_p$ is the classical radius of proton, $\tau_y$ is the transverse damping time and $T_{0}$ is the revolution time.

\section{Machine parameters choice}

The design goal of energy of SPPC is about 70-100TeV using the same tunnel with CEPC which is about 50Km. A larger circumference like 100Km for SPPC is also being considered. We want to use the superconducting magnets which is about 20T \citep{lab10}. We can develop a systematic way to calculate the parameter starting from the maximum beam beam tune shift limit and the design goal. Our design goal is: luminosity $L_0$, beam energy $E_0$, ring circumference $C_0$ and IP numbers $N_{IP}$. Table 1 shows the goals, known quantities and constants.

\begin{center}
\tabcaption{ \label{tab1}  The design goal and known quantities.}
\begin{tabular}{|c|c|}
\hline
Circumference &  $C_0 =54.7Km$  \\ \hline
Beam Energy   &  $E_0 =35TeV$   \\ \hline
IP numbers    &  $N_{IP}=2$     \\ \hline
Luminosity    &  $L = 1.0 \times 10^{35}cm^{-2}s^{-1}$   \\ \hline
Total straight section length  & $L_{SS}=7595m$ \\ \hline
Arc filling factor   & $f_1=0.79$ \\ \hline
Bunch filling factor & $f_2=0.80$  \\ \hline
Energy gain($15\thicksim20$) & $Gain=16.67$ \\ \hline
Total/inelastic cross section  &  $\sigma_{cross}=140mbarn$ \\ \hline
Light spead  & $c=3 \times 10^8 m/s$  \\ \hline

\end{tabular}
\end{center}

The luminosity for pp collider can be writen as \citep{lab4}:

\begin{equation}
\mathcal{L}=\frac{I_b}{e} \frac{\xi_y}{\beta^*} \frac{\gamma}{r_p} F_{ca} F_h
\end{equation}

\begin{equation}
\mathcal{L}_0=\frac{I_b}{e} \frac{\xi_y}{\beta^*} \frac{\gamma}{r_p}
\end{equation}

Where, $F_{ca}$ is the luminosity reduction factor due to cross angle \citep{lab11}:
\begin{equation}
F_{ca}=\frac{1}{\sqrt{1+(\frac{\sigma_z \theta_c}{2\sigma^*})^2}}
\end{equation}

$F_h$ is the luminosity reduction factor due to hourglass effect \citep{lab12}:
\begin{equation}
F_h=\frac{\beta^*}{\sqrt{\pi}\sigma_z} \exp{(\frac{{\beta^*}^2}{2\sigma_z^2})} K_0(\frac{{\beta^*}^2}{2\sigma_z^2})
\end{equation}

Put $\xi_{y,max}$ into the luminosity formula, we can get:

\begin{equation}
\mathcal{L}_0=\frac{I_b}{e} \frac{\xi_{y,max}}{\beta^*} \frac{\gamma}{r_p}=\frac{2845}{2\pi r_p e f(x)} \frac{1}{\beta^*} \sqrt{\frac{I_b P_{SR} \gamma}{2 E_0 N_{IP}}}
\end{equation}

And, then the beta function at IP can be written as:

\begin{equation}
\beta^*=\frac{2845}{2\pi r_p e f(x)} \frac{1}{\mathcal{L}_0} \sqrt{\frac{I_b P_{SR} \gamma}{2 E_0 N_{IP}}}
\end{equation}

The RMS IP spot size:($\sigma^*=\sigma_x=\sigma_y$)

\begin{equation}
\sigma^*=\sqrt{\beta^* \epsilon}=\sqrt{\beta^*\frac{\epsilon_n}{\gamma}}
\end{equation}

Beta at the 1st parasitic encounter with bunch separation $\Delta t$ :

\begin{equation}
l_1 =c\times \Delta t
\end{equation}

\begin{equation}
\beta_1 =\beta^*+\frac{(l_1/2)^2}{\beta^*}
\end{equation}

RMS spot size at the 1st parasitic encounter:

\begin{equation}
\sigma_1=\sqrt{\beta_1 \epsilon}=\sqrt{\beta_1\frac{\epsilon_n}{\gamma}}
\end{equation}

The full cross angle \citep{lab4}:

\begin{equation}
\theta_c=\frac{2 \times 6 \sigma_1}{l_1 /2}=\frac{24\sigma_1}{l_1}
\end{equation}

We can rewrite $F_{ca}$ as:

\begin{equation}
F_{ca}=\frac{1}{\sqrt{1+\Phi^2}}
\end{equation}

\begin{equation}
\begin{split}
\Phi &=\frac{\sigma_z \theta_c}{2\sigma^*}=\frac{12\sigma_z \sigma_1}{l_1 \sigma^*}=\frac{12\sigma_z \sqrt{\beta_1 \frac{\epsilon_n}{\gamma}}}{l_1
\sqrt{\beta^* \frac{\epsilon_n}{\gamma}}}=\frac{12\sigma_z}{l_1} \sqrt{\frac{\beta_1}{\beta^*}} \\
&=12 \sqrt{\frac{\sigma_z^2}{(c\Delta t)^2}+\frac{1}{4(\beta^*/\sigma_z)^2}}
\end{split}
\end{equation}

Where $\Phi$ is Piwinski angle, $\beta^*$ is beta function at IP, $\sigma_z$ is bunch length and $\Delta t$ is the bunch separation.

When the luminosity reduce less than $10\%$ due to the crossing angle effect, we have $F_{ca}\geqslant0.9$. From equation(17) we get :

\begin{equation}
\Phi \leqslant 0.434822 (rad)
\end{equation}

Bunch numbers:
\begin{equation}
n_b=\frac{T_0 f_2}{\Delta t}
\end{equation}

Bunch population:
\begin{equation}
N_p=\frac{I_b}{n_b f_{rev} e}
\end{equation}

Combining equation(11)(18)(19)(20)(21), we can get reasonable values of $\beta^*$ $I_b$ $\Delta t$ $n_b$ $N_p$ and the ratio $\beta^*/\sigma_z$, where should also consider the instability influence and the constraints from technic.

From the definition of beam beam tune shift \citep{lab11}:
\begin{equation}
\xi_y=\frac{N_p r_p}{4\pi \epsilon_n}
\end{equation}

We can get the normalized emittance:
\begin{equation}
\epsilon_n=\frac{N_p r_p}{4\pi \xi_{y,max} }
\end{equation}

Then we can calculate $\sigma^*$ $\beta_1$ $\sigma_1$ $\theta_c$ and $F_h$. Finally, we get the final value of the luminosity:

\begin{equation}
\mathcal{L}=\mathcal{L}_0 F_{ca} F_h
\end{equation}

We can also calculate the follow parameters easily.

Energy loss per turn \citep{lab13}:

\begin{equation}
U_0=0.00778\left[ MeV \right] \frac{(E_0 \left[TeV\right])^4}{\rho \left[m \right]}
\end{equation}

SR power per ring:

\begin{equation}
P_{SR}=U_0 I_b
\end{equation}

Critical photon energy$\left[E_c\right]$ \citep{lab13}\citep{lab14}:

\begin{equation}
E_c \left[KeV \right]=1.077 \times 10^{-4} (E_0 \left[TeV\right])^2 B\left[ T \right]
\end{equation}

Accumulated particles per beam:

\begin{equation}
N_{ACC}=N_p n_b
\end{equation}

Stored energy per beam:

\begin{equation}
W=N_{ACC} E_0 e = N_p n_b E_0 e
\end{equation}

ARC SR heat load \citep{lab15}:

\begin{equation}
\text{SR heat load}=\frac{P_{SR}}{L_{Dipole}}
\end{equation}

Transverse damping time$\left[\tau_x\right]$ \citep{lab16}:

\begin{equation}
\tau_x= \frac{2 E_0 T_0}{J_x U_0}
\end{equation}

Longitudinal damping time$\left[\tau_\varepsilon\right]$ \citep{lab16}:

\begin{equation}
\tau_\varepsilon= \frac{2 E_0 T_0}{J_\varepsilon U_0}
\end{equation}

Beam life time due to burn-off \citep{lab11}:

\begin{equation}
\tau_{burn-off}=\frac{N_p n_b}{\mathcal{L} N_{IP} \sigma_{cross}}=\frac{N_{ACC}}{\mathcal{L} N_{IP} \sigma_{cross}}
\end{equation}

The time required to reach $1/e$ of the initial luminosity \citep{lab11}:

\begin{equation}
\tau_{1/e}=(\sqrt{e}-1)\times \tau_{burn-off}
\end{equation}

Other contributions to luminosity decay come from Toucheck scattering and from particle losses due to a slow emittance blow-up. An emittance blow-up can be caused by the scattering of particles on the residual gas, the nonlinear force of the beam-beam interaction, RF noise and IBS scattering effects. The synchrotron radiation damping decreases the bunch dimensions and can partially compensate the beam size blow-up due to the above effects. Assuming that the radiation damping process just cancels the beam blow up due to the beam-beam interactions and RF noise, one can estimate the net luminosity lifetime by \citep{lab11}:

\begin{equation}
\tau_L=\frac{1}{\frac{1}{\tau_{IBS}}+\frac{2}{\tau_{rest-gas}}+\frac{1}{\tau_{1/e}}}
\end{equation}

If the run time $\tau_{run}$ fulfils equation(36), the integrated luminosity has the maximum value and the run time will be the optimum run time \citep{lab11}.

\begin{equation}
\log(\frac{\tau_{turn-around}+\tau_{run}}{\tau_L}+1)=\frac{\tau_{run}}{\tau_L}
\end{equation}

\begin{equation}
\tau_{optimum}=\tau_{run}
\end{equation}

Integrating the luminosity over one luminosity run$\left[fb^{-1}\right]$:

\begin{equation}
L_{int}=\mathcal{L}\tau_L (1-e^\frac{-\tau_{run}}{\tau_L})\times\frac{3600}{10^{39}}
\end{equation}

where $\tau_{run}$ is the optimum total length of the luminosity run.

The overall collider efficiency depends on the ratio of the run length and the average turnaround time. So the optimum average integrated luminosity/day$\left[fb^{-1}\right]$ is \citep{lab11}:

\begin{equation}
L_{tot}=\frac{24}{\tau_{run}[h]+\tau_{turn-around}[h]} L_{int}
\end{equation}

\vspace{5mm}

As a summary, we obtain a set of machine parameters with luminosity goal $L_0$, beam energy $E_0$, ring circumference $C_0$ and IP numbers $N_{IP}$.

\begin{equation}
U_0=0.00778\left[ MeV \right] \frac{(E_0 \left[TeV\right])^4}{\rho \left[m \right]}
\end{equation}

\begin{equation}
E_c \left[KeV \right]=1.077 \times 10^{-4} (E_0 \left[TeV\right])^2 B\left[ T \right]
\end{equation}

\begin{equation}
P_{SR}=U_0 I_b
\end{equation}

\begin{equation}
\xi_{y,max}= \frac{2845\gamma}{f(x)}\sqrt{\frac{r_p}{6\pi R N_{IP}}}=\frac{2845}{2\pi f(x)}\sqrt{\frac{T_0}{\tau_y \gamma N_{IP}}}=\frac{\xi_1}{f(x)}
\end{equation}

\begin{equation}
f(x) = 1- \frac{2}{ \sqrt{2\pi}}\int_0^x \mathrm{e}^{{-}\frac{t^2}{2}} \,\mathrm{d}t
\end{equation}

\begin{equation}
x^2=\frac{4f(x)}{\pi \xi_{y,max} N_{IP}}=\frac{4f(x)^2}{\pi \xi_1 N_{IP}}
\end{equation}

\begin{equation}
\mathcal{L}_0=\frac{I_b}{e} \frac{\xi_{y,max}}{\beta^*} \frac{\gamma}{r_p}=\frac{2845}{2\pi r_p e f(x)} \frac{1}{\beta^*} \sqrt{\frac{I_b P_{SR} \gamma}{2 E_0 N_{IP}}}
\end{equation}

\begin{equation}
\beta^*=\frac{2845}{2\pi r_p e f(x)} \frac{1}{\mathcal{L}_0} \sqrt{\frac{I_b P_{SR} \gamma}{2 E_0 N_{IP}}}
\end{equation}

\begin{equation}
\sigma^*=\sqrt{\beta^* \epsilon}=\sqrt{\beta^*\frac{\epsilon_n}{\gamma}}
\end{equation}

\begin{equation}
l_1 =c\times \Delta t
\end{equation}

\begin{equation}
\beta_1 =\beta^*+\frac{(l_1/2)^2}{\beta^*}
\end{equation}

\begin{equation}
\sigma_1=\sqrt{\beta_1 \epsilon}=\sqrt{\beta_1\frac{\epsilon_n}{\gamma}}
\end{equation}

\begin{equation}
\theta_c=\frac{2 \times 6 \sigma_1}{l_1 /2}=\frac{24\sigma_1}{l_1}
\end{equation}

\begin{equation}
F_{ca} =\frac{1}{\sqrt{1+(\frac{\sigma_z \theta_c}{2\sigma^*})^2}} =\frac{1}{\sqrt{1+\Phi^2}}
\end{equation}

\begin{equation}
\Phi=\frac{\sigma_z \theta_c}{2\sigma^*} =12 \sqrt{\frac{\sigma_z^2}{(c\Delta t)^2}+\frac{1}{4(\beta^*/\sigma_z)^2}}
\end{equation}

\begin{equation}
n_b=\frac{T_0 f_2}{\Delta t}
\end{equation}

\begin{equation}
N_p=\frac{I_b}{n_b f_{rev} e}
\end{equation}

\begin{equation}
\epsilon_n=\frac{N_p r_p}{4\pi \xi_{y,max}}
\end{equation}

\begin{equation}
F_h=\frac{\beta^*}{\sqrt{\pi}\sigma_z} \exp{(\frac{{\beta^*}^2}{2\sigma_z^2})} K_0(\frac{{\beta^*}^2}{2\sigma_z^2})
\end{equation}

\begin{equation}
\mathcal{L}=\mathcal{L}_0 F_{ca} F_h
\end{equation}

\begin{equation}
N_{ACC}=N_p n_b
\end{equation}

\begin{equation}
W=N_{ACC} E_0 e = N_p n_b E_0 e
\end{equation}

\begin{equation}
\text{SR heat load}=\frac{P_{SR}}{L_{Dipole}}
\end{equation}

\begin{equation}
\tau_x= \frac{2 E_0 T_0}{J_x U_0}
\end{equation}

\begin{equation}
\tau_\varepsilon= \frac{2 E_0 T_0}{J_\varepsilon U_0}
\end{equation}

\begin{equation}
\tau_{burn-off}=\frac{N_{ACC}}{\mathcal{L} N_{IP} \sigma_{cross}}
\end{equation}

\begin{equation}
\tau_{1/e}=(\sqrt{e}-1)\times \tau_{burn-off}
\end{equation}

\begin{equation}
\tau_L=\frac{1}{\frac{1}{\tau_{IBS}}+\frac{2}{\tau_{rest-gas}}+\frac{1}{\tau_{1/e}}}
\end{equation}

\begin{equation}
\log(\frac{\tau_{turn-around}+\tau_{run}}{\tau_L}+1)=\frac{\tau_{run}}{\tau_L}
\end{equation}

\begin{equation}
\tau_{optimum}=\tau_{run}
\end{equation}

\begin{equation}
L_{int}=\mathcal{L}\tau_L (1-e^\frac{-\tau_{run}}{\tau_L})\times\frac{3600}{10^{39}}
\end{equation}

\begin{equation}
L_{tot}=\frac{24}{\tau_{run}[h]+\tau_{turn-around}[h]} L_{int}
\end{equation}

\section{Compare the LHC parameter list with the parameter obtained by our method}

To check our method, we use it to chose and calculate the LHC parameters and compare them with the LHC parameter list\citep{lab15}. The second column in Table 2 is the parameter obtained using our systematical method, which is reasonable and nearly with the parameters in LHC parameter list. This indicates that our method is reasonable and more powerful. We can use this method to design and choose parameters for any proton proton circular colliders.

\begin{center}
\tabcaption{ \label{tab1} Compare the LHC parameter list with the parameter obtained by our method.}
\footnotesize

\begin{tabular}
{|p{120pt}|p{28pt}|p{25pt}|p{20pt}|}
\hline
\raisebox{-1.50ex}[0cm][0cm]{}&
LHC-list&
LHC-new&
 \\
\cline{2-4}
 &
\multicolumn{2}{|p{70pt}|}{Value} &
Unit \\
\hline
\multicolumn{4}{|p{190pt}|}{\textbf{Main parameters and geometrical aspects}}  \\
\hline
Beam energy[$E_0$]&
7&
7&
TeV \\
\hline
Circumference[$C_0$]&
26.7&
26.7&
km \\
\hline
Lorentz gamma[$\gamma $]&
7463&
7463&
 \\
\hline
Dipole field[B]&
8.33&
8.26&
T \\
\hline
Dipole curvature radius[\textit{$\rho $}]&
2801&
2826&
m \\
\hline
Bunch filling factor[f2]&
0.78&
0.80&
 \\
\hline
Arc filling factor[f1]&
0.79&
0.79&
 \\
\hline
Total dipole magnet length[L$_{Dipole}$]&
17599&
17756&
m \\
\hline
Arc length[L$_{ARC}$]&
22476&
22476&
m \\
\hline
Total straight section length[Lss]&
4224&
4224&
m \\
\hline
Energy gain factor in collider rings&
15.6&
15.6&
 \\
\hline
Injection energy [E$_{inj}$]&
0.45&
0.45&
TeV \\
\hline
Number of IPs[N$_{IP}$]&
4&
2&
 \\
\hline
\multicolumn{4}{|p{190pt}|}{\textbf{Physics performance and beam parameters}}  \\
\hline
Peak luminosity per IP[L]&
1.0E+34&
1.0E+34&
$/cm^{2}s$ \\
\hline
Optimum run time&
15.2&
10.46&
hour \\
\hline
Optimum average integrated luminosity/day&
0.47&
0.42&
$fb^{-1}$ \\
\hline
Assumed turnaround time&
6&
5&
hour \\
\hline
Overall operation cycle&
21.2&
16.0&
hour \\
\hline
Beam life time due to burn-off[$\tau $]&
45&
40.65&
hour \\
\hline
Total / inelastic cross section[$\sigma $]&
111/85&
111/85&
mbarn  \\
\hline

\multicolumn{4}{|p{190pt}|}{\textbf{Beam parameters}}  \\

\hline
Beta function at collision[\textit{$\beta $*}]&
0.55&
0.56&
m \\
\hline
Max beam-beam tune shift perIP[\textit{$\xi $y}]&
0.0033&
0.0032&
 \\
\hline
Number of IPs contributing to $\Delta $Q&
3&
2&
 \\
\hline
Max total beam-beam tune shift&
0.01&
0.0064&
 \\
\hline
Circulating beam current[I$_{b}$] &
0.584&
0.589&
A \\
\hline
Bunch separation[\textit{$\Delta $t}]&
25 \par 5&
25 \par 5&
ns \\
\hline
Number of bunches[n$_{b}$]&
2808 \par &
2848&
 \\
\hline
Bunch population[Np]&
1.15 \par &
1.15&
$10^{11}$ \\
\hline

\end{tabular}

\end{center}

\begin{center}
\footnotesize

\begin{tabular}
{|p{120pt}|p{25pt}|p{25pt}|p{20pt}|}

\hline

Normalized RMS transverse emittance[$\varepsilon $]&
3.75 \par &
4.39&
$\mu $m \\
\hline
RMS IP spot size[$\sigma $*]&
16.7&
16.09&
$\mu $m \\
\hline
Beta at the 1st parasitic encounter[\textit{$\beta $}1]&
26.12&
32.37&
m \\
\hline
RMS spot size at the 1st parasitic encounter[$\sigma _{1}$]&
114.6&
138&
$\mu $m \\
\hline
RMS bunch length[$\sigma $z]&
75.5&
75.7&
mm \\
\hline
Accumulated particles per beam&
0.32&
0.33&
$10^{15}$ \\
\hline
Full crossing angle[$\theta $c]&
285&
441.16&
$\mu $rad \\
\hline
Reduction factor according to cross angle[Fca]&
0.8391&
0.7788&
 \\
\hline
Reduction factor according to hour glass effect[Fh]&
0.9954&
0.9956&
 \\
\hline
\multicolumn{4}{|p{190pt}|}{\textbf{Other beam and machine parameters}}  \\
\hline
Energy loss per turn[U$_{0}$]&
0.0067&
0.0066&
MeV \\
\hline
Critical photon energy[Ec]&
0.044&
0.044&
KeV \\
\hline
SR power per ring[P$_{0}$]&
0.0036&
0.0039&
MW \\
\hline
Stored energy per beam[W]&
0.362&
0.367&
GJ \\
\hline
RF voltage[$V_{rf}$]&
16&
16&
MV \\
\hline
RF Frequency[$f_{rf}$]&
400.8&
400.8&
MHz \\

\hline
Revolution frequency[$f_{rev}$]&
11.236&
11.236&
kHz \\
\hline

Harmonic number&
35671&
35671&
 \\
\hline
rms energy spread[$\delta_\epsilon$]&
1.129&
1.124&
$10^{-4}$\\
\hline
Momentum compaction factor [$\alpha_p$]&
3.225&
3.26&
$10^{-4}$\\
\hline
Synchrotron tune[$\nu_s$]&
1.904&
2.057&

$10^{-3}$\\
\hline
Synchrotron Frequency[fsyn]&

21.4&
23.12&
Hz \\
\hline
Bucket area&
8.7&
9.4&
eVs \\
\hline
Bucket half height($\Delta $E/E)&
0.36&
0.35&
$10^{-3}$ \\
\hline
Arc SR heat load per aperture&
0.206&
0.22&
W/m \\
\hline
Damping partition number [Jx]&
1&
1&
 \\
\hline
Damping partition number [Jy]&
1&
1&
 \\
\hline
Damping partition number [J$\varepsilon $]&
2&
2&
 \\
\hline
Transverse damping time [$\tau $x]&
25.8&
26.18&
hour \\
\hline
Longitudinal damping time [$\tau \varepsilon $]&
12.9&
13.09&
hour \\
\hline
\end{tabular}

\end{center}

\section{Parameter choice for SPPC}

\subsection{Parameter scan}

Using the method above, we scan the goal luminosity $\mathcal{L}_0$ with different bending radius $\rho$, IP numbers $N_{IP}$ and different ratio of  $\beta^*/\sigma_z$. Table 3 shows the input parameters. We get some meaningful results which are shown From Fig.1 to Fig.8.

\begin{center}
\tabcaption{ \label{tab3}  Input parameters for machine design.}

\footnotesize

\begin{tabular}
{|p{65pt}|p{70pt}|p{85pt}|}

\hline
{Energy $E_0$} & {Circumference $C_0$} & {Goal luminosity $\mathcal{L}_0$}  \\ \hline
$35.0TeV$ & $54.7Km$ & $(1\thicksim4)\times 10^{35}cm^{-2}s^{-1}$ \\ \hline
{IP numbers $N_{IP}$} & {Bending radius $\rho$} & {ratio of $ \beta^*/\sigma_z$}\\ \hline
$2\thicksim4$ & $5.9\thicksim6.5Km$ & $10\thicksim20$ \\ \hline
\end{tabular}
\end{center}

\begin{center}
 \includegraphics[width=8cm]{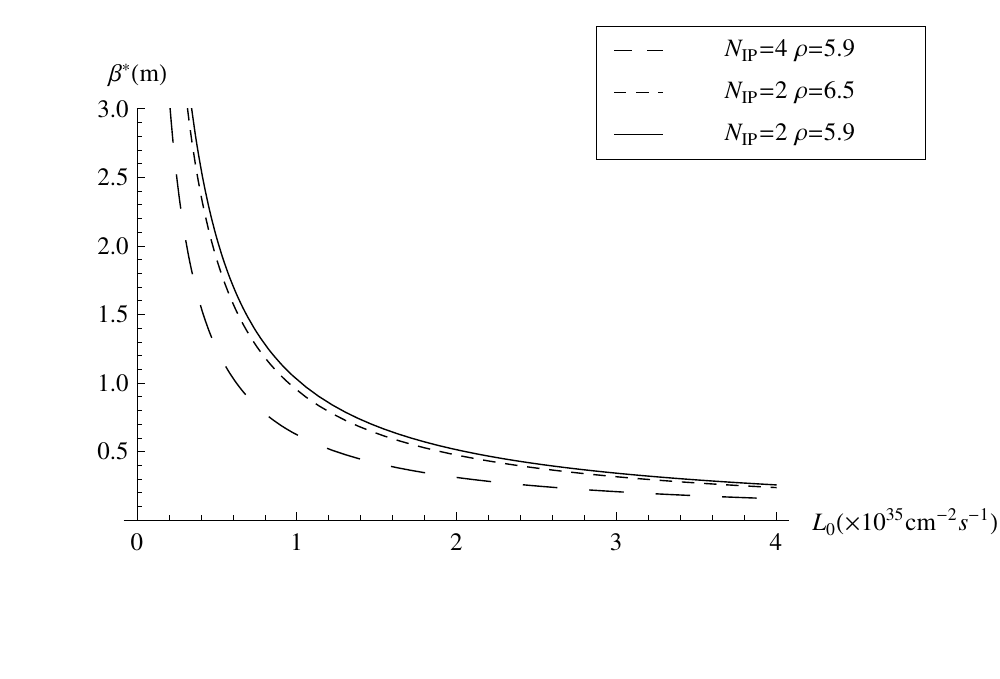}
\figcaption{\label{fig1}  Vertical beta at IP as the function of goal luminosity. }
\end{center}

Fig.1 shows that larger luminosity needs smaller vertical IP beta function. Larger bending radius and more interaction points require smaller $\beta^*$ at the same goal luminosity.

\begin{center}
  \includegraphics[width=8cm]{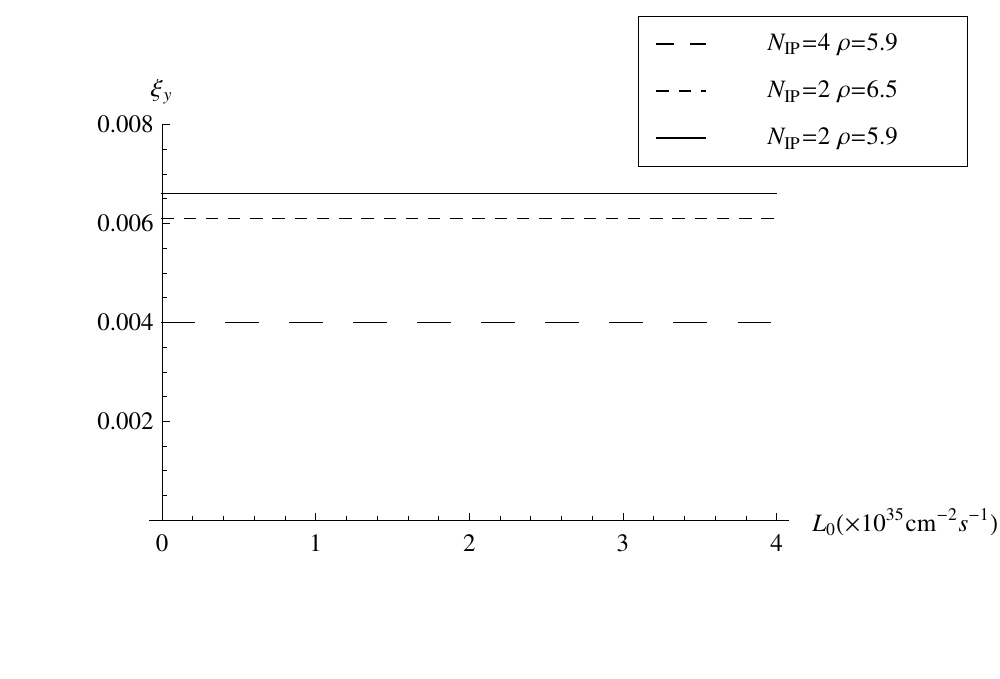}
\figcaption{\label{fig1}  Vertical beam beam tune shift as the function of peak luminosity. }
\end{center}

Fig.2 shows smaller bending radius and less interaction points give larger vertical beam-beam tune shift.

\begin{center}
  \includegraphics[width=8cm]{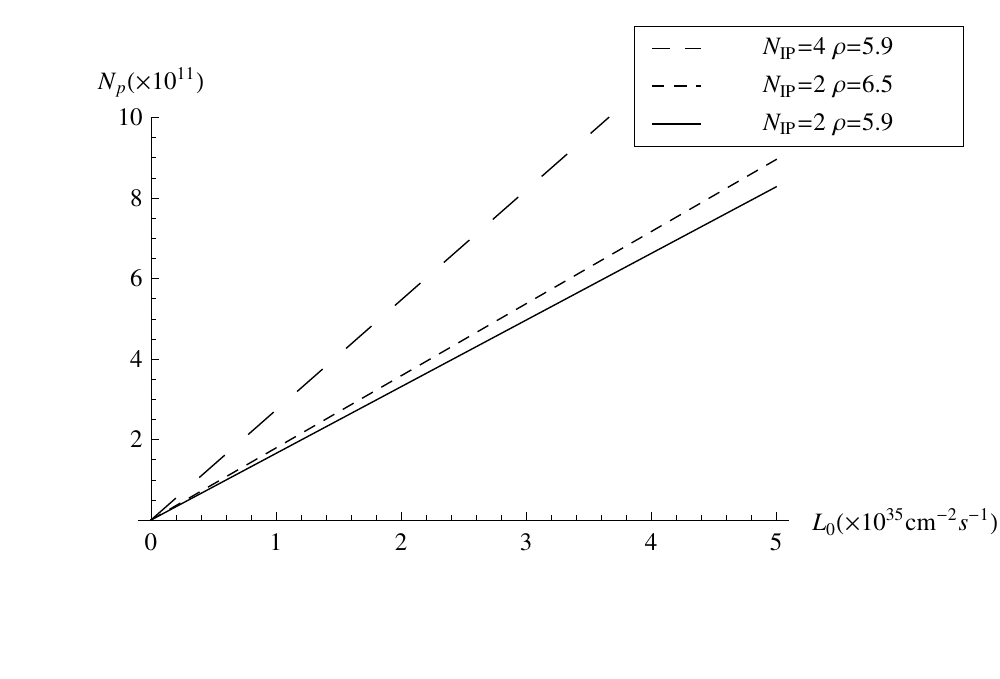}
\figcaption{\label{fig3}  Bunch population as the function of peak luminosity. }
\end{center}

\begin{center}
  \includegraphics[width=8cm]{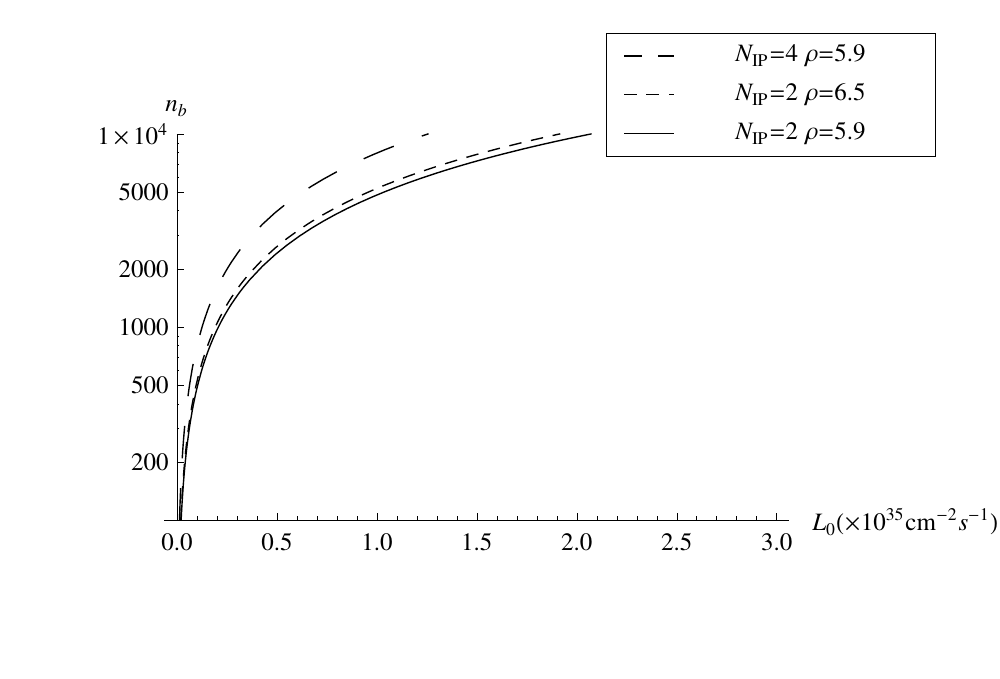}
\figcaption{\label{fig3}  Bunch number as the function of peak luminosity. }
\end{center}

Fig.3 and Fig.4 show that larger luminosity needs larger bunch population or larger bunch number. Larger bending radius and more interaction points indicate larger bunch population or larger bunch number at the same goal luminosity.

\begin{center}
  \includegraphics[width=8cm]{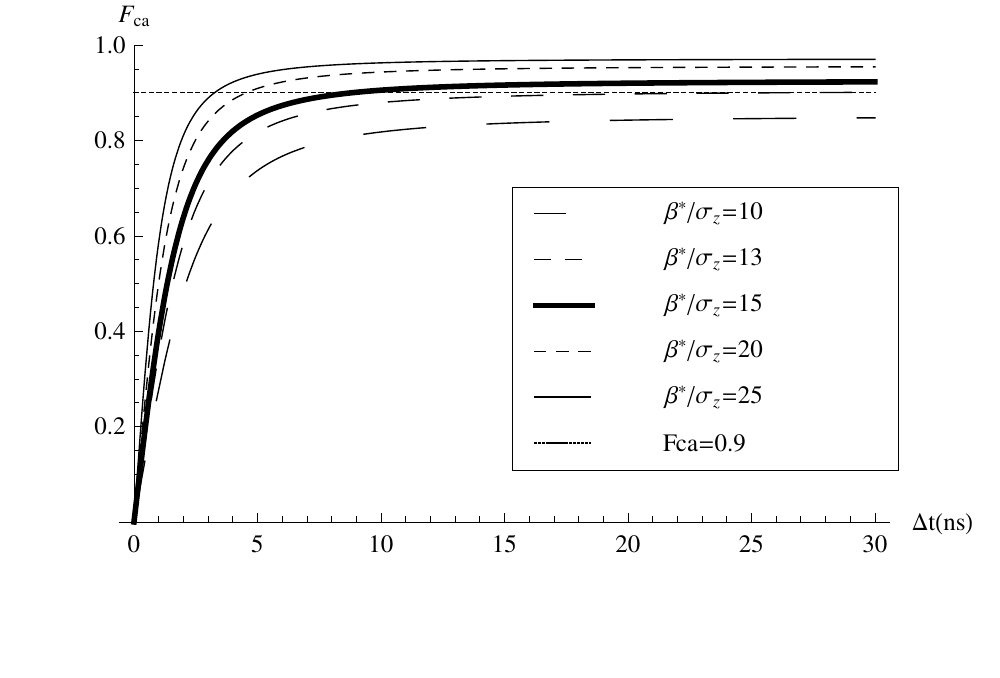}
\figcaption{\label{fig5}  $F_{ca}$ as the function of $\Delta t$. }
\end{center}

\begin{center}
  \includegraphics[width=8cm]{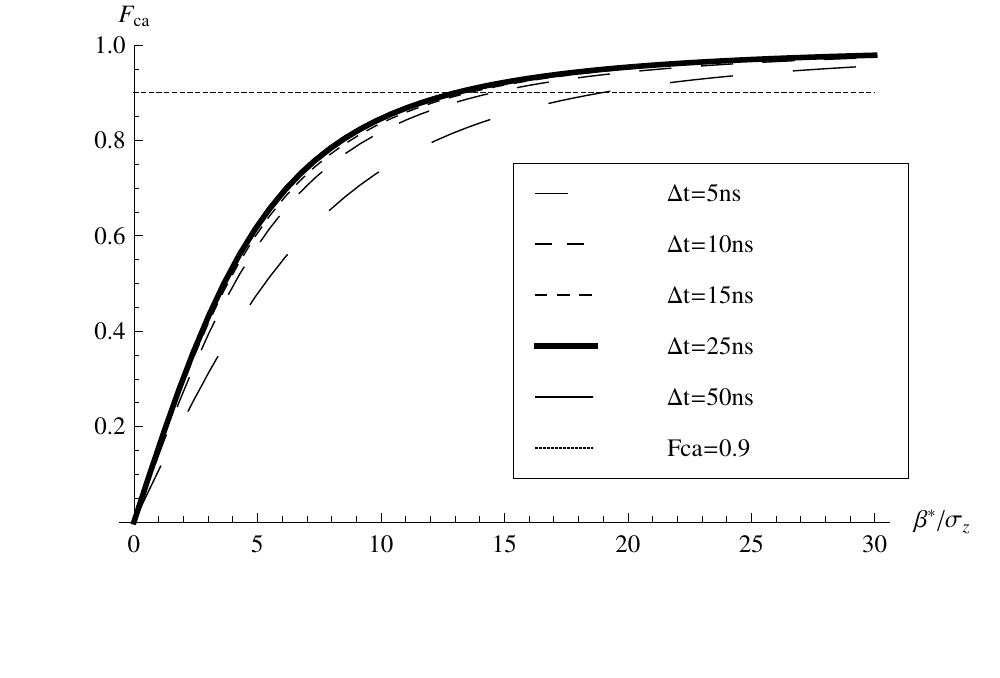}
\figcaption{\label{fig6}  $F_{ca}$ as the function of the ratio of $\beta^*$  and $\sigma_z$. }
\end{center}

Fig.5 and Fig.6 tell us that the reduction factor according to cross angle has relationship with bunch separation($\Delta t$) and the ratio of IP beta and RMS bunch length($\beta^*/\sigma_z$). The maximum value of this factor is 1, and larger $\beta^*/\sigma_z$ makes this valve nearer to 1. If we want this effect reduce the luminosity less than 10\%, we should have $F_{ca} \geqslant 0.9$. The dashed line in Fig.5 and Fig.6 is the value equal to 0.9, and we can easily get the important information from the figures. We should choose a larger $\beta^*/\sigma_z$, which about 15 is much reasonable and now the bunch separation is $25ns$. If we want to choose a smaller bunch separation like $5ns$, the ratio of $\beta^*$ and $\sigma_z$ should be more than 20. We should consider both of them and choose the eclectic values. Fig.7 shows the 3D diagram of the relationship of $F_{ca}$, $\Delta t$ and $\beta^*/\sigma_z$.

\begin{center}
  \includegraphics[width=8cm]{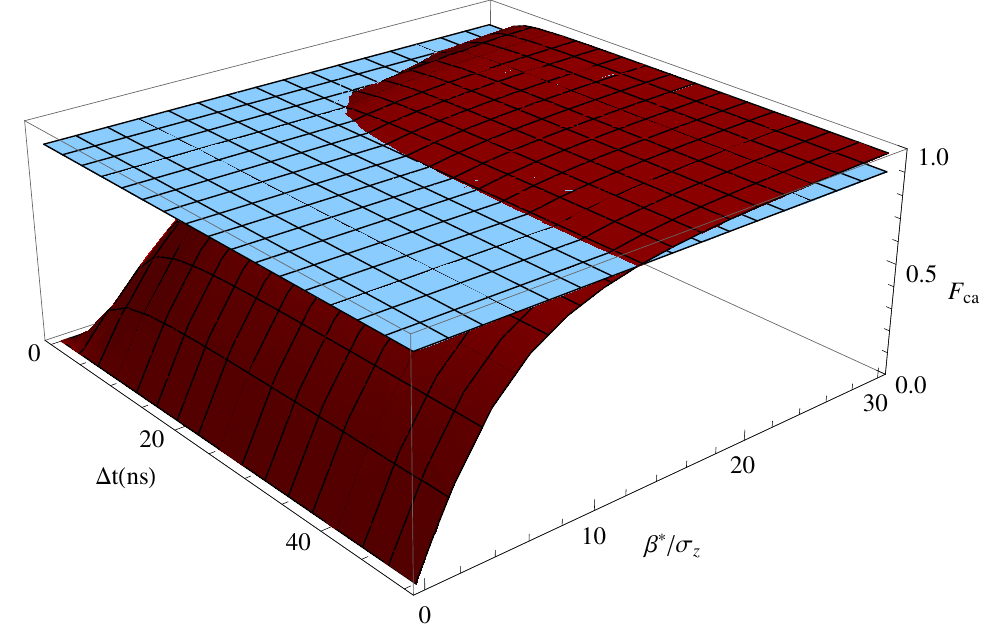}
\figcaption{\label{fig7}  The 3D diagram of the relationship of $F_{ca}$ $\Delta t$ and $\beta^*/\sigma_z$. }
\end{center}

\begin{center}
  \includegraphics[width=8cm]{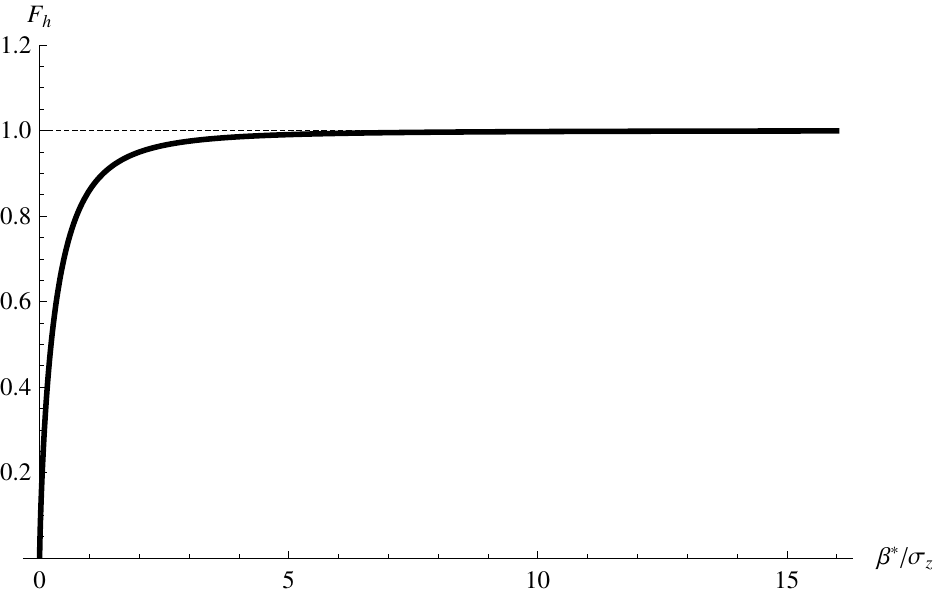}
\figcaption{\label{fig8}   $F_h$ as the function of the ratio of $\beta^*$  and $\sigma_z$. }
\end{center}

Fig.8 shows the reduction factor according to hourglass effect as the function of the ratio of IP $\beta$ function and RMS bunch length. A large ratio makes larger $F_h$ value. To reduce the reduction of luminosity according to hourglass effect, we should also choose a reasonable larger ratio of $\beta^*$ and $\sigma_z$.

Overall speaking, we should decrease IP numbers and increase bending radius in order to achieve higher luminosity. $N_{IP}=2$ is the reasonable minimum value for IP number. Assuming the maximum dipole arc filling factor is 80\%, 5.9 km bending radius will be a limit for the 54.7 km ring.

\subsection{Constraints form RF system}

As long as a set of beam parameters is determined, we need to check the RF system to see if the bunch length can be achieved. Firstly, considering the synchrotron radiation energy loss has to be compensated by the RF cavities, one finds \citep{lab16}:

\begin{equation}
U_0=e V_{rf} \sin(\phi_s)
\end{equation}

where $V_{rf}$ is the total voltage of the RF cavities and $\phi_s$ is the synchrotron phase. According to eq. (72), one gets

\begin{equation}
\phi_s=\pi-\arcsin(\frac{U_0}{e V_{rf}})
\end{equation}

We can estimate the RF frequency from the "pill-box" model. As the following picture shows. We can find the $f_{rf}$ via the Maxwell's equation and the boundary conditions \citep{lab16}.

\begin{equation}
J_0(\frac{\omega}{c}R_0)=0
\end{equation}

\begin{equation}
\frac{\omega}{c}R_0=2.405\quad
\frac{2\pi f_{rf}}{c}R_0=2.405
\end{equation}

\begin{equation}
f_{rf}=\frac{2.405 c}{2\pi R_0}
\end{equation}

When the cavity inner radius $R_0=30cm$,$f_{RF}=400 MHz$ is a reasonable choose \citep{lab16}.

\begin{center}
 \includegraphics[width=8cm]{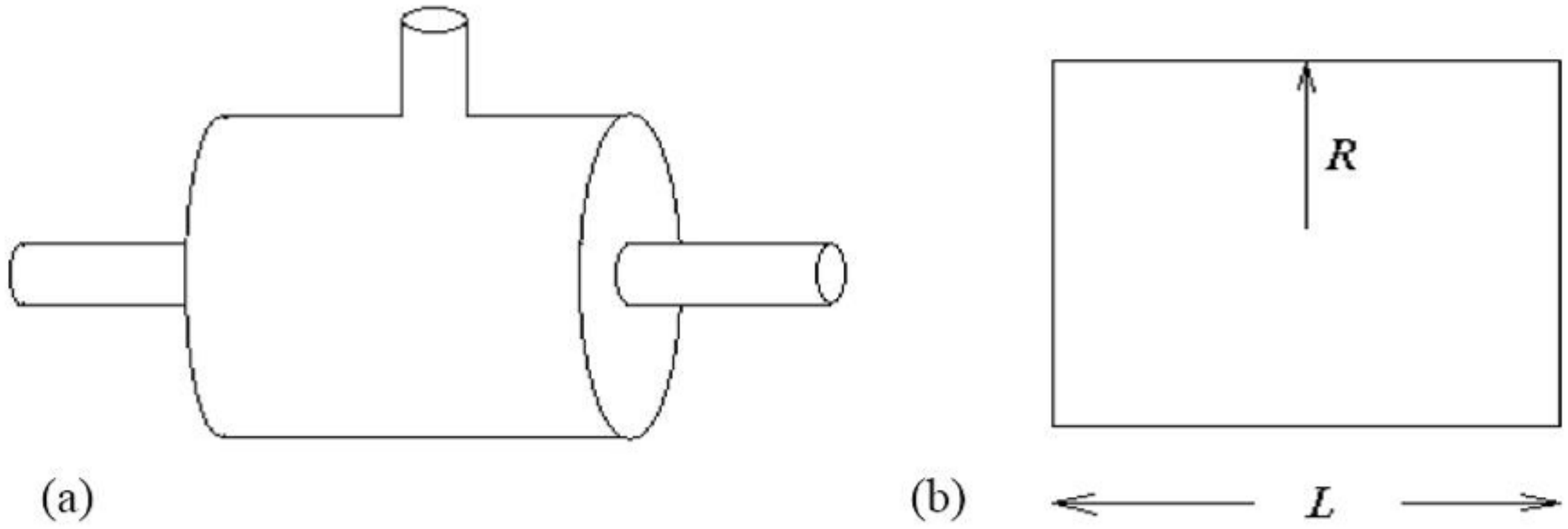}
\figcaption{\label{fig9}  Pill-box model. }
\end{center}

In a storage ring with an isomagnetic guide field (one which has a constant radius $\rho$ in the magnets and is straight elsewhere) the relative energy spread $\sigma_\epsilon/E_0$ can be expressed as \citep{lab17}:

\begin{equation}
(\delta_\epsilon)^2=(\frac{\sigma_\epsilon}{E_0})^2=\frac{C_q \gamma^2}{J_\epsilon \rho_0}  \quad (isomag)
\end{equation}

so,

\begin{equation}
\delta_\epsilon=\gamma \sqrt{\frac{C_q }{J_\epsilon \rho_0}}
\end{equation}

where $C_q=1.2817 \times 10^{-12} m$ is a constant.

The nature bunch length is expressed by \citep{lab17}:

\begin{equation}
\sigma_l=\frac{\alpha_p R \delta_\epsilon}{\nu_s}
\end{equation}

where, $\alpha_p$ is the momentum compaction factor, $R$ is the average radius of the ring. $\nu_s$ is the longitudinal oscillation tune which can be expressed as:

\begin{equation}
\nu_s =\sqrt{-\frac{\eta_p h e V_{rf} \cos\phi_s}{2 \pi E_s \beta_s^2}}
\end{equation}

Where $\eta_p$ is the phase slippage factor, when $v\approx c$,\quad $\beta \approx 1$ ,\quad $\gamma \rightarrow \infty$, \quad $\eta_p=\alpha_p-\frac{1}{\gamma^2}\approx\alpha_p$, and $ h= f_{rf}/f_{rev}=f_{rf} T_0$, we can rewrite $\nu_s$ as follow:

\begin{equation}
\nu_s =\sqrt{-\frac{\alpha_p f_{rf} T_0 e V_{rf} \cos\phi_s}{2 \pi E_0}}
\end{equation}

And then the nature bunch length can be expressed as \citep{lab17}\citep{lab18}:

\begin{equation}
\sigma_l =\sqrt{-\frac{2 \pi E_0\alpha_p }{f_{rf} T_0 e V_{rf} \cos\phi_s}}R \delta_\epsilon
\end{equation}

The energy acceptance can be expressed as \citep{lab17}\citep{lab18}:

\begin{equation}
\eta_{acceptance}=\sqrt{\frac{2 U_0}{\pi \alpha_p f_{rf} T_0 E_0 }[\sqrt{q^2-1}-\arccos(\frac{1}{q})]}
\end{equation}

where, $q= e V_{rf}/U_0$. Combining the eqs.(82) and eqs.(83), we can get the RF frequency $f_{rf}$ and the momentum compaction $\alpha_p$ for given RF voltage $ V_{rf}$ and energy acceptance $\eta$.

The synchrotron frequency \citep{lab18}:

\begin{equation}
f_{syn}=\frac{\nu_s}{T_0}=\nu_s f_{rev}
\end{equation}

Bucket area \citep{lab18}:

\begin{equation}
\text{bucket area}=\frac{16 \nu_s}{h|\eta_p|\sqrt{|\cos{\phi_s}|}}\alpha(\phi_s)
\end{equation}
where the dimensionless function $\alpha(\phi_s)$is the bucket area normalized to the case when $\phi_s=0$. For the case $\eta_p < 0$, we have

\begin{equation}
\alpha ( \phi_s )= \frac{1}{4\sqrt{2}} \int\nolimits_{\phi_2}^{\pi-\phi_s}[\cos\phi+\cos{\phi_s}-(\pi-\pi_s)\sin{\phi_s}]^{1/2}{\rm d} \phi
\end{equation}
when $\phi_s=0$, $\alpha(\phi_s)=1$, then the  bucket area is $\frac{16 \nu_s}{h|\eta_p|}$.

The bucket half height \citep{lab18}:

\begin{equation}
\text{bucket half height}=\sqrt{\frac{2 e V_0 |\cos{\phi_s}-\frac{\pi-2 \phi_s}{2}\sin{\phi_s}|}{\pi \beta_s^2 E_s h |\eta_p|}}
\end{equation}
when $\phi_s=0$ ,we have bucket half height $\frac{2 \nu_s}{h|\eta_p|}$.

\subsection{Machine parameter choice for SPPC}

Combining the discussions above, we get a set of new design for the 54.7 km SPPC. We also tried to give a set of parameters for larger circumference SPPC, like 78Km or 100Km. Table 4 is the parameter list for SPPC. As a comparation, we put the parameter for LHC HL-LHC HE-LHC and FCC-hh together in Table 4.\citep{lab4}\citep{lab10} The first plan for SPPC is using the same tunnel with CEPC. The circumference is 54.7Km which is determined by CEPC. We choose the dipole field as 20T and get center-of-mass energy 70TeV. If we want to explore the higher energy, we should make the circumference larger. When we want to explore center-of-mass energy 100TeV and keep the dipole field 20T, the circumference should be 78Km at least. At this condition, there is hardly space to upgrade. So a 100Km SPPC is much better because the dipole field is only 14.7T at this condition. If we make the dipole field 20T too, we can get the center-of-mass energy as high as 136TeV.

\end{multicols}

\begin{center}
\tabcaption{ \label{tab4}  Parameter lists for LHC HL-LHC HE-LHC FCC-hh and SPPC.}
\footnotesize
\begin{tabular}
{|p{116pt}|p{28pt}|p{28pt}|p{28pt}|p{30pt}|p{33pt}|p{28pt}|p{28pt}|p{28pt}|p{28pt}|p{30pt}|}
\hline
\raisebox{-1.50ex}[0cm][0cm]{}&
LHC&
HL-LHC&
HE-LHC&
FCC-hh&
SPPC-Pre-CDR&
SPPC-54.7Km&
SPPC-100Km&
SPPC-100Km&
SPPC-78Km&
 \\
\cline{2-11}
 &
\multicolumn{9}{|p{275pt}|}{Value} &
Unit \\
\hline
\multicolumn{11}{|p{423pt}|}{\textbf{Main parameters and geometrical aspects}}  \\
\hline
Beam energy[$E_0$]&
7&
7&
16.5&
50&
35.6 &
35.0&
50.0 &
68.0&
50.0 &
TeV \\
\hline
Circumference[$C_0$]&
26.7&
26.7&
26.7&
100(83)&
54.7&
54.7&
100&
100&
78&
km \\
\hline
Lorentz gamma[$\gamma $]&
7463&
7463&
14392&
53305&
37942 &
37313 &
53305 &
72495 &
53305 &
 \\
\hline
Dipole field[B]&
8.33&
8.33&
20&
16(20)&
20&
19.69&
14.73&
20.03&
19.49&
T \\
\hline
Dipole curvature radius[\textit{$\rho $}]&
2801&
2801&
2250&
10416 \par (8333.3)&
5928&
5922.6 &
11315.9 &
11315.9 &
8549.8&
m \\
\hline

\end{tabular}
\end{center}

\begin{center}
\footnotesize
\begin{tabular}
{|p{118pt}|p{28pt}|p{28pt}|p{28pt}|p{30pt}|p{30pt}|p{28pt}|p{30pt}|p{30pt}|p{30pt}|p{30pt}|}
\hline
Bunch filling factor[$f_2$]&
0.78&
0.78&
0.63&
0.79&
0.8&
0.8 &
0.8 &
0.8 &
0.8 &
 \\
\hline
Arc filling factor[$f_1$]&
0.79&
0.79&
0.79&
0.79&
0.79&
0.79&
0.79&
0.79&
0.79&
 \\
\hline
Total dipole magnet length [L$_{Dipole}$]&
17599&
17599&
14062&
65412 \par (52333)&
37246&
37213&
71100&
71100&
53720&
m \\
\hline
Arc length[L$_{ARC}$]&
22476&
22476&
22476&
83200 \par (66200)&
47146&
47105&
90000&
90000&
68000&
m \\
\hline
Total straight section length[$L_{ss}$]&
4224&
4224&
4224&
16800&
7554&
7595&
10000&
10000&
10000&
m \\
\hline
Energy gain factor in collider rings&
15.6&
15.6&
13.5&
15.2&
17.0&
16.67&
17.5&
17.5&
17.5&
 \\
\hline
Injection energy [E$_{inj}$]&
0.45&
0.45&
$>$1.0&
3.3&
2.1 &
2.1 &
2.9 &
3.9 &
2.9 &
TeV \\
\hline
Number of IPs[N$_{IP}$]&
4&
2&
2&
2&
2&
2&
2&
2&
2&
 \\
\hline
\multicolumn{11}{|p{423pt}|}{\textbf{Physics performance and beam parameters}}  \\
\hline
Peak luminosity per IP[L]&
1.0E+34&
5.0E+34&
5.0E+34&
5.0E+34&
1.2E+35&
1.2E+35&
1.52E+35&
1.02E+36&
1.52E+35&
cm$^{ - 2}$s$^{ - 1}$ \\
\hline
Optimum run time&
15.2&
10.2&
5.8&
12.1(10.7)&
5.87&
5.87&
6.69&
2.47&
5.91&
hour \\
\hline
Optimum average integrated luminosity/day&
0.47&
2.8&
1.4&
2.2(2.1)&
3.36&
3.36&
4.84&
12.97&
4.28&
$fb^{-1}$ \\

\hline
Assumed turnaround time&
6&
&
&
5&
5&
5&
5&
5&
5&
hour \\
\hline
Overall operation cycle&
21.2&
&
&
17.4(16.3)&
11.5&
11.5&
12.5&
8.0&
12.0&
hour \\
\hline
Beam life time due to burn-off[$\tau $]&
45&
15.4&
5.7&
19.1(15.9)&
9.65&
9.65&
12.74 &
2.07&
9.78&
hour \\
\hline
Total / inelastic cross section[$\sigma $]&
111/85&
111/85&
129/93&
153/108&
140&
140&
155&
160&
155&
mbarn  \\

\hline
\multicolumn{11}{|p{423pt}|}{\textbf{Beam parameters}}  \\
\hline
Beta function at collision[\textit{$\beta $*}]&
0.55&
0.15 \par (min)&
0.35&
1.1&
0.75&
0.85&
0.97&
0.24&
1.06&
m \\
\hline
Max beam-beam tune shift perIP[\textit{$\xi $y}]&
0.0033&
0.0075&
0.005&
0.005&
0.006&
0.0065&
0.0067&
0.008&
0.0073&
 \\
\hline
Number of IPs contributing to $\Delta $Q&
3&
2&
2&
2&
2&
2&
2&
2&
2&
 \\
\hline
Max total beam-beam tune shift&
0.01&
0.015&
0.01&
0.01&
0.012&
0.013&
0.0134&
0.016&
0.0146&
 \\
\hline
Circulating beam current[I$_{b}$] &
0.584&
1.12&
0.478&
0.5&
1.0&
1.024 &
1.024 &
1.024 &
1.024 &
A \\
\hline
Bunch separation[\textit{$\Delta $t}]&
25 \par 5&
25 \par 5&
25 \par 5&
25 \par 5&
25&
25&
25&
25&
25&
ns \\
\hline
Number of bunches[n$_{b}$]&
2808 \par &
2808&
2808&
10600 \par (8900) \par 53000 \par (44500)&
5835&
5835 &
10667 &
10667 &
8320&
 \\
\hline
Bunch population[Np]&
1.15 \par &
2.2&
1&
1.0 \par 0.2&
2.0&
2.0&
2.0&
2.0&
2.0&
$10^{11}$ \\

\hline
Normalized RMS transverse emittance[$\varepsilon $]&
3.75 \par &
2.5&
1.38&
2.2 \par 0.44&
4.10&
3.72&
3.65 &
3.05&
3.36&
$\mu $m \\
\hline
RMS IP spot size[$\sigma $*]&
16.7&
7.1&
5.2&
6.8&
9.0&
8.85 &
7.85 &
3.04&
7.86&
$\mu $m \\
\hline
Beta at the 1st parasitic encounter[\textit{$\beta $}1]&
26.12&
93.9&
40.53&
13.88&
19.5&
18.70&
16.51&
64.1&
15.36&

m \\
\hline
RMS spot size at the 1st parasitic encounter[$\sigma _{1}$]&
114.6&
177.4&
62.3&
23.9&
45.9&
43.2 &
33.6&
51.9&
31.14&
$\mu $m \\
\hline
RMS bunch length[$\sigma $z]&
75.5&
75.5&
75.5&
80(75.5)&
75.5&
56.5&
65&
15.8&
70.6&
mm \\
\hline
Accumulated particles per beam&
0.32&
0.62&
0.28&
1.06(0.89) \par 5.3(4.45)&
1.2&
1.17&
2.13&
2.13&
1.66&
$10^{15}$ \\
\hline
Full crossing angle[$\theta $c]&
285&
590&
185&
74&
73&
138 &
108 &
166&
99&
$\mu $rad \\
\hline
Reduction factor according to cross angle[Fca]&
0.8391&
0.314&
0.608&
0.910&
0.8514&
0.9257 &
0.9248 &
0.9283&
0.9248&
 \\
\hline
Reduction factor according to hour glass effect[Fh]&
0.9954&
0.9491&
0.9889&
0.9987&
0.9975&
0.9989&
0.9989&
0.9989&
0.9989&
 \\
\hline
\multicolumn{11}{|p{423pt}|}{\textbf{Other beam and machine parameters}}  \\
\hline
Energy loss per turn[U$_{0}$]&
0.0067&
0.0067&
0.201&
4.6(5.86)&
2.10&
1.97 &
4.30 &
14.7&
5.69&
MeV \\
\hline
Critical photon energy[Ec]&
0.044&
0.044&
0.575&
4.3(5.5)&
2.73&
2.60&
3.97&
9.96&
5.25&
KeV \\
\hline
SR power per ring[P$_{0}$]&
0.0036&
0.0073&
0.0962&
2.4(2.9)&
2.1&
2.0 &
4.4 &
15.1&
5.82&
MW \\
\hline
Stored energy per beam[W]&
0.362&
0.694&
0.701&
8.4(7.0)&
6.6&
6.53 &
17.1 &
23.21&
13.31&
GJ \\
\hline

\end{tabular}

\end{center}

\begin{center}
\footnotesize
\begin{tabular}
{|p{118pt}|p{28pt}|p{28pt}|p{28pt}|p{30pt}|p{30pt}|p{28pt}|p{30pt}|p{30pt}|p{30pt}|p{30pt}|}
\hline
RF voltage[V$_{rf}$]&
16&
16&
16&
16&
16&
16&
16&
16&
16&
MV \\
\hline
RF Frequency[$f_{rf}$]&
400.8&
400.8&
400.8&
400.8&
400.8&
400.8&
400.8&
400.8&
400.8&
MHz \\

\hline
Revolution frequency[$f_{rev}$]&
11.236&
11.236&
11.236&
3.00&
5.48&
5.48 &
3.00 &
3.00 &
3.84&
kHz \\
\hline

Harmonic number&
35671&
35671&
35671&
133600&
73079&
73079&
133600&
133600&
104208&
 \\
\hline

rms energy spread[$\delta_\epsilon$]&
1.129&
1.13&
2.97&
4.17&
3.9&
3.88&
4.01&
5.46&
4.6&
$10^{-4}$\\
\hline
Momentum compaction factor [$\alpha_p$]&
$3.225\times10^{-4}$&
$3.92\times10^{-4}$&
$2.23\times10^{-5}$&
$1.48\times10^{-6}$&
$3.39\times10^{-6}$&
$1.48\times10^{-6}$&
$6.79\times10^{-7}$&
$6.56\times10^{-9}$&
$7.54\times10^{-7}$&
\\
\hline
Synchrotron tune[$\nu_s$]&
1.904&
2.26&
0.35&
0.098&
0.133&
0.088&
0.067&
0.0036&
0.061&
$10^{-3}$\\
\hline
Synchrotron Frequency[fsyn]&
21.4&
25.33&
3.93&
0.29&
0.73&
0.48&
0.20&
0.011&
0.24&
Hz \\
\hline
Bucket area&
8.7&
8.6&
23.51&
27.01&
28.63&
43.74&
39.94&
348.55&
42.94&
eVs \\
\hline
Bucket half height($\Delta $E/E)&
0.36&
0.32&
0.87&
0.78&
0.96&
1.48&
1.17&
1.94&
1.14&
$10^{-3}$ \\
\hline
Arc SR heat load per aperture&
0.206&
0.33&
4.35&
28.4(44.3)&
57.8&
54.24 &
61.9 &
211.72&
108.41&
W/m \\
\hline

Damping partition number [Jx]&
1&
1&
1&
1&
1&
1&
1&
1&
1&
 \\
\hline
Damping partition number [Jy]&
1&
1&
1&
1&
1&
1&
1&
1&
1&
 \\
\hline
Damping partition number [J$\varepsilon $]&
2&
2&
2&
2&
2&
2&
2&
2&
2&
 \\
\hline
Transverse damping time [$\tau $x]&
25.8&
25.8&
2.0&
1.08(0.64)&
1.71&
1.80&
2.15&
0.86&
1.27&
hour \\
\hline
Longitudinal damping time [$\tau \varepsilon $]&
12.9&
12.9&
1.0&
0.54(0.32)&
0.85&
0.90&
1.08&
0.43&
0.635&
hour \\
\hline
\end{tabular}

\end{center}

\begin{multicols}{2}

\section{Comparing beam-beam tune shift of SPPC with LHC HL-LHC HE-LHC and FCC-hh}

In the parameter design of LHC HL-LHC HE-LHC and FCC-hh, the CERN people assume the beam-beam tune shift limit as a constant number \citep{lab4}\citep{lab11}. But we can find the beam-beam parameter has relationship with several parameters. A method to estimate the maximum beam-beam tune shift limit was developed in refrernce \citep{lab9}. We compare the calculated numbers with the parameter list chosen numbers and find that these calculated numbers by analytical expression are much reasonable according to the real experimental numbers. We can easily get the ratio of the beam-beam tune shift of the list chosen number and the calculated number. The result was shown in Table 5, we can find that HL-LHC's choice is much overlarge and the other machines' choices are more reasonable.

\end{multicols}

\begin{center}
\tabcaption{ \label{tab6}  Comparing beam-beam tune shift of SPPC with LHC HL-LHC HE-LHC and FCC-hh.}
\footnotesize
\begin{tabular}
{|p{138pt}|p{25pt}|p{25pt}|p{28pt}|p{28pt}|p{28pt}|p{30pt}|p{30pt}|p{26pt}|p{28pt}|}
\hline
&
LHC \par 7TeV&
HL-LHC \par 7TeV&
HE-LHC \par 16.5TeV&
FCC-hh \par 50TeV&
SPPC-Pre-CDR\par 35.6TeV&
SPPC-54.7Km \par 35TeV&
SPPC-100Km \par 50TeV&
SPPC-100Km \par 68TeV&
SPPC-78Km \par 50TeV \\
\hline
Number of IPs contributing to $\Delta $Q&
3&
2&
2&
2&
2&
2&
2&
2&
2 \\
\hline
Max total beam-beam tune shift&
0.01&
0.015&
0.01&
0.01&
0.012&
0.013&
0.0134&
0.016&
0.0146\\
\hline
Max beam-beam tune shift perIP [\textit{$\xi $y}] (parameter list)&
0.0033&
0.0075&
0.005&
0.005&
0.006&
0.0065&
0.0067&
0.008&
0.0073 \\
\hline
Max beam-beam tune shift perIP [\textit{$\xi $y}] (calculated)&
0.00321&
0.00321&
0.00499&
0.00685&
0.00662&
0.006559&
0.006688&
0.00801&
0.00731\\
\hline
[\textit{$\xi $y}](parameter list)/ [\textit{$\xi $y}](calculated)&
1.0287&
2.3379&
1.002&
0.7299&
0.9063&
0.9910&
1.001&
0.9986&
0.9999 \\
\hline
\end{tabular}

\end{center}

\vspace{5mm}
\begin{multicols}{2}

\section{Conclusion}

In this paper, a systematic method of how to make an appropriate parameter choice for a circular pp collider by using analytical expression of beam-beam tune shift limit started from given luminosity goal, beam energy and technical limitations was developed. By using this method, we reveal the relations of machine parameters with goal luminosity clearly and hence give a parameter choice in an efficient way. We also show the parameter chose for a 50Km SPPC and larger circumference SPPC, like a 78Km SPPC or a 100Km SPPC.

\end{multicols}

\vspace{10mm}
\centerline{\rule{80mm}{0.1pt}}
\vspace{2mm}

\begin{multicols}{2}

\end{multicols}

\clearpage

\begin{thebibliography}{90}

\vspace{3mm}

\bibitem{lab1}Frank Zimmermann. HE-LHC {\&} VHE-LHC accelerator overview, injector chain, and main parameter choices. Joint Snowmass-EuCARD/AccNet-HiLumi LHC meeting Frontier Capabilities for Hadron Colliders 2013, 21 February 2013.

\bibitem{lab2}Frank Zimmermann. CERN Future Circular Colliders Study. International Workshop on Future High Energy Circular Colliders. IHEP Beijing, 16 December 2013.

\bibitem{lab3}Daniel Schulte. FCC-hh. FCC-hh kick-off meeting, March 2014.

\bibitem{lab4}Future Circular Collider Study Hadron Collider Parameters. fcc-coordination-group, fcc-leaders, FCC-ACC-SPC-0001, 2014-02-11.


\bibitem{lab5}D.Wang, J.Gao. Study on Beijing Higgs Factory(BHF) and Beijing Hadron Collider(BHC). IHEP-AC-LC-Note2012-012.

\bibitem{lab6}Alain Blondel, Alex Chao, Weiren Chou et al. Accelerators for a Higgs Factory: Linear vs. Circular. Reports of the ICFA Beam Dynamics Workshop, HF2012

\bibitem{lab7}Yifang Wang. Introduction of CEPC-SppC. Geneve, Feb. 13, 2014.

\bibitem{lab8}J. Gao. Emittance growth and beam lifetime limitations due to beam-beam effects in e+e- storage rings. Nucl. Instr. and methods A533£¨2004£©p. 270-274.

\bibitem{lab9}J. Gao, M. Xiao, F. Su et al. Analytical Estimation of Maximum Beam-beam Tune Shifts for Electron-Postron and Hadron Circular Colliders. HF2014,Beijing,Oct.9-12,2014.

\bibitem{lab10}Tang Jingyu, Zou Ye, Su Feng et al. SPPC-Pre-CDR. 2014.10, p.3-6.

\bibitem{lab11}LHC Design Report. European Organization for Nuclear Research, 2004. Volume 1, Chapter3, p21-22.

\bibitem{lab12}D. Wang, J. Gao, M. Xiao et al. Optimization Parameter Design of a Circular e+e- Higgs Factory. IHEP-AC-LC-Note2013-005, ILC-ÎïÀí-2013-03, Feb. 22nd, 2013

\bibitem{lab13}Zhang Yuan. Accelerator Physics. Lectures of UCAS 2012. Chapter9-10.

\bibitem{lab14}William A. Barletta. Unit 11 - Lecture 18 :Synchrotron Radiation - I. US Particle Accelerator School 2008

\bibitem{lab15}LHC Design Report. European Organization for Nuclear Research, 2004. Volume 1, Chapter2, p3-13.

\bibitem{lab16}Qin Qing. Accelerator Physics. Lectures of UCAS 2012. P48-57.

\bibitem{lab17}M.Sands. The Physics of Electron Storage Rings. SLAC-121-UC-28(ACC). p.113-129.

\bibitem{lab18}Zhang Yuan. Accelerator Physics. Lectures of UCAS 2012. Chapter8-9.




\end{thebibliography}
\end{document}